\documentclass[twocolumn,aps,superscriptaddress,showpacs,floatfix,prc]{revtex4}
\usepackage{mathrsfs}
\usepackage{amssymb}
\usepackage{amsmath}
\usepackage{graphicx}
\usepackage[normalem]{ulem}
\usepackage[dvips]{color}
\usepackage{bm}
\usepackage{longtable}

\renewcommand{\v}[1]{\textbf{#1}}
\renewcommand{\rm}[1]{\textrm{#1}}
\renewcommand{\d}{\mathrm{d}}

\def\esym{$E_{sym}(\rho)$~}

\def\es0{$E_{sym}(\rho_0)$~}

\def\us0{$U_{sym}(\rho_0,k_F)$~}

\def\l0{$L(\rho_0)$~}

\begin{document}

\title{Isospin dependence of nucleon effective masses in neutron-rich matter}

\author{Bao-An Li}
\affiliation{Department of Physics and Astronomy, Texas A$\&$M
University-Commerce, Commerce, TX 75429-3011, USA}

\author{Bao-Jun Cai}
\affiliation{Department of Physics and Astronomy, Texas A$\&$M
University-Commerce, Commerce, TX 75429-3011, USA}

\author{Lie-Wen Chen }
\affiliation{Department of Physics and Astronomy and Shanghai Key Laboratory for Particle Physics and Cosmology, Shanghai Jiao Tong University, Shanghai 200240, China}

\author{Xiao-Hua Li}
\affiliation{School of Nuclear Science and Technology, University of South China, Hengyang 421001, China}

\begin{abstract}
In this talk, we first briefly review the isospin dependence of the total nucleon effective mass $M^{\ast}_{J}$ inferred from analyzing nucleon-nucleus scattering data within an isospin dependent non-relativistic optical potential model, and the isospin dependence of the nucleon E-mass $M^{\ast,\rm{E}}_{J}$ obtained from applying the Migdal-Luttinger theorem to a phenomenological single-nucleon momentum distribution in nuclei constrained by recent electron-nucleus scattering experiments. Combining information about the isospin dependence of both the nucleon total effective mass and E-mass, we then infer the isospin dependence of nucleon k-mass using the well-known relation $M^{\ast}_{J}=M^{\ast,\rm{E}}_{J}\cdot M^{\ast,\rm{k}}_{J}$. Implications of the results on the nucleon mean free path (MFP) in neutron-rich matter are discussed. 
\end{abstract}


\maketitle

\section{Nucleon effective masses}\label{sec.I}
To ease the following discussions, we first recall the basic definitions and relations of the three distinct nucleon effective masses used typically in non-relativistic descriptions of nuclear matter and give a few examples of model predictions.

The k-mass $M^{\ast,\rm{k}}_{J}$ and E-mass $M^{\ast,\rm{E}}_{J}$ of a nucleon $J=n/p$ characterizes, respectively, the space and time non-locality of nuclear interactions. They are normally obtained from the momentum and energy dependence of the single-nucleon potential $U_{J} (\rho,\delta, k,E)$ in nucleonic matter of density $\rho$ and isospin asymmetry $\delta\equiv (\rho_n-\rho_p)/\rho$ via \cite{Jeu76,Mah85,jamo}
\begin{equation}
\frac{M^{\ast,\rm{E}}_{J}}{M}=1-\frac{\partial U_{J}}{\partial E}~\rm{and}~
\frac{M^{\ast,\rm{k}}_{J}}{M}=\left[1+\frac{M_{J}}{|\v{k}|}\frac{\partial U_{J}}{\partial|\v{k}|}\right]^{-1}
\end{equation}
where $M$ is the average mass of nucleons in free-space.
Once an energy-momentum dispersion relation is assumed using the on-shell condition $E=k^2/2M+U_{J} (\rho,\delta, k,E)$, an equivalent single-particle potential either local in space or time can be obtained. The so-called total effective mass $M^{\ast}_{J}$
\begin{eqnarray}\label{em1}
\frac{M^{*}_{J}}{M}&=&1-\frac{\d U_{J}(\v{k}(E),E,\rho,\delta)}{dE}\Bigg|_{E(k_{\rm{F}}^{J})}\\\nonumber
&=&\left[1+\frac{M}{\hbar^2k_{\rm{F}}^J}\frac{\d U_{J}(\v{k},E(\v{k}),\rho,\delta)}{\d|\v{k}|}\Bigg|_{k_{\rm{F}}^J}\right]^{-1}
\end{eqnarray}
is then used to characterize equivalently either the momentum or energy dependence of the single-nucleon potential. We emphasize that once nucleons are put on shell, the total effective mass is the only effective mass one can extract from either the first or the second part of the above equation. As we shall discuss later, one then has to use other approaches to evaluate the E-mass and k-mass.
The total effective mass is a measure of the energy level density. The  well-known relationship
\begin{equation}\label{METK}
M^{\ast}_{J}=M^{\ast,\rm{E}}_{J}\cdot M^{\ast,\rm{k}}_{J}
\end{equation}
among the three kinds of nucleon effective masses can be derived by noticing that \cite{Jeu76}
\begin{equation}
\frac{dE}{dk}\equiv \frac{k}{M_J^*}=\frac{k}{M}+\frac{\partial U}{\partial k}+\frac{\partial U}{\partial E}\cdot \frac{dE}{dk}.
\end{equation}
In the above, $k_{\rm{F}}^J=(1+\tau_3^J\delta)^{1/3}\cdot k_{\rm{F}}$ with
$k_{\rm{F}}=(3\pi^2\rho/2)^{1/3}$  being the nucleon Fermi momentum
in symmetric nuclear matter at density $\rho$, $\tau_3^{J}=+1$ or $-1$ for neutrons or protons.
\begin{figure}[htb]
%\vspace{0.8cm}
\includegraphics[scale=0.9,width=8cm]{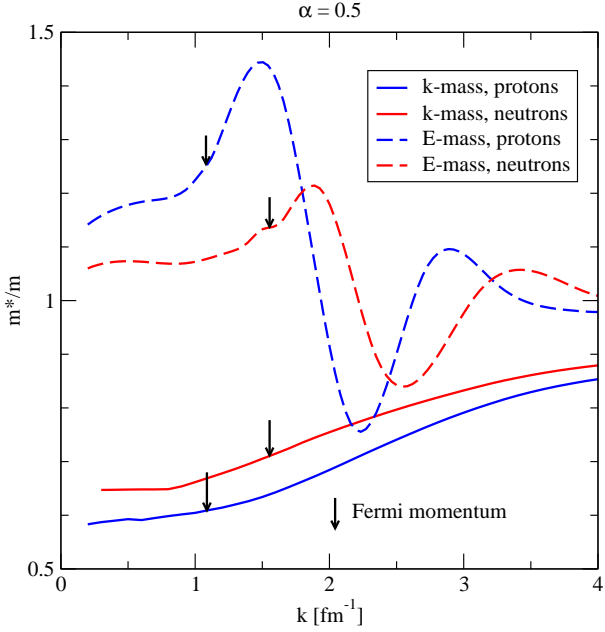}
\caption{Effective k-masses (solid lines) and E-masses (dashed lines) of neutrons (red) and protons (blue) derived from the BHF self-energies using the CD-Bonn interactions for nucleonic matter with an isospin asymmetry of 0.5 at saturation density. Taken from Ref. \cite{mu04} of Hassaneen and M\"uther.}
\label{Muther}
\end{figure}

\begin{figure}[htb]
%\includegraphics[scale=0.6]{Fig2-Baldo.pdf}
%\vspace{-1.2cm}
\includegraphics[scale=0.8,width=\columnwidth]{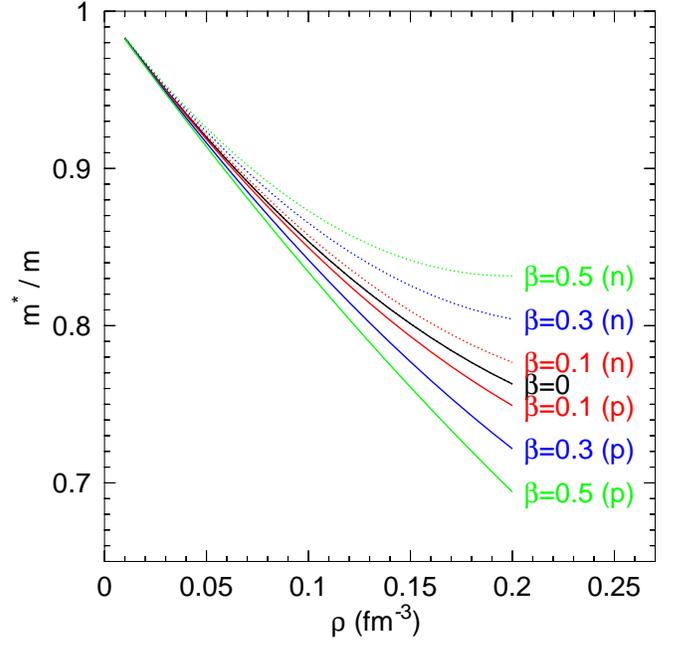}
\caption{Proton (p, full line) and neutron (n, dotted line) total effective masses as a function
of density for different values of the isospin asymmetry parameter $\beta$ from a BHF calculation by Baldo et al. in Ref.  \cite{Baldo16}. }
\label{Baldo}
\end{figure}

Many microscopic many-body theories using various interactions have been used in calculating all three kinds of nucleonic effective masses, see, e.g., ref. \cite{LiChen15} for a recent review.
Shown in Fig. \ref{Muther} and Fig. \ref{Baldo} are examples of Brueckner-Hartree-Fock (BHF) predictions of the E-mass, k-mass and total effective mass of neutrons and protons in asymmetric
nuclear matter using some of the most widely used nuclear interactions. The recent focus of many studies has been on the splitting of the neutron-proton effective masses and its dependence on the isospin asymmetry and density of the neutron-rich medium encountered in heavy-ion collisions and in some astrophysical situations \cite{mrs,page06}, such as in neutron stars and neutrino spheres of supernova explosions. A thorough understanding of the nucleon effective masses is critical for us to better understand many interesting issues in both nuclear physics and astrophysics. Generally, most of the models predict that in neutron-rich medium, neutrons have a k-mass and total
effective mass higher than those for protons, and protons have a higher E-mass than neutrons at their respective Fermi surfaces. However, depending on the models and interactions used, the predictions
can change dramatically. For example, some of the widely used Skyrme interactions predict that protons have a higher total effective mass than neutrons in neutron-rich matter. Thus, it is very important to extract reliable information about the nucleon effective masses from experiments \cite{Jun15,Lynch16}. While conclusions from recent analyses of heavy-ion experiments using transport models are still quite ambiguous even about the sign of the neutron-proton total effective mass splitting \cite{Jun15,Lynch16}, it is very encouraging that analyses of nucleon-nucleus and electron-nucleus scatterings can constrain clearly at least the sign of the neutron-proton total and E-mass splitting, respectively, at saturation densities \cite{XHLi,CaiLi16a}. In this talk, we shall briefly review theses results and then infer from them the neutron-proton k-mass splitting at saturation density.
For more details, please see the original publications in Refs. \cite{XHLi,CaiLi16a,LiBA13}.

\section{Relation between the neutron-proton effective mass splitting and symmetry energy in neutron-rich matter}\label{sec.II}
Assuming the energy on-shell condition has been used, then the single-nucleon potential can be written as a function of momentum k, i.e., $U_{J}(k,\rho,\delta)$. The latter is the well-known Lane potential that can be expanded as \cite{Lan62}
\begin{equation}\label{sp}
U_{J}(k,\rho,\delta)=U_0(k,\rho)+\tau_3 U_{sym}(k,\rho)\cdot\delta+\mathcal{O}(\delta^2),
\end{equation}
where $U_0(k,\rho)$ and $U_{sym}(k,\rho)$ are the isoscalar and isovector potential, respectively.
The neutron-proton effective mass splitting
$m^*_{n-p}(\rho,\delta)\equiv(M_{\rm n}^*-M_{\rm p}^*)/M$ can be written as \cite{LiBA13}
\begin{align}\label{Em2}
m^*_{n-p}=\frac{\frac{M}{\hbar^2}\left(\frac{1}{k_F^p}\frac{dU_p}{dk}\mid_{k_F^p}-\frac{1}{k_F^n}\frac{dU_n}{dk}\mid_{k_F^n}\right)}{\left[1+\frac{M_p}{\hbar^2k_F^p}\frac{dU_p}{dk}\mid_{k_F^p}\right]\left[1+\frac{M_n}{\hbar^2k_F^n}\frac{dU_n}{dk}\mid_{k_F^n}\right]}.
\end{align}
Up to the first-order in isospin asymmetry parameter $\delta$, the above expression can be further simplified to
\begin{equation}\label{npe1}
m^*_{n-p}\approx 2\delta\frac{M}{\hbar^2k_F}\left[-\frac{dU_{sym}}{dk}-\frac{k_F}{3}\frac{d^2U_0}{dk^2}+\frac{1}{3}\frac{dU_0}{dk}\right]_{k_F}\left(\frac{M^*_0}{M}\right)^2.
\end{equation}
Generally, the $m^*_{n-p}$ depends on the momentum dependence of  both the isovector $U_{sym}$ and isoscalar $U_0$ potentials. Interestingly, the same factors also determine the density dependence of nuclear symmetry energy. The latter is currently the most uncertain part of the equation of state of neutron-rich matter and has significant implications for many areas of both nuclear physics and astrophysics, see, e.g., ref. \cite{EPJA} for a recent review.  Using the Hugenholtz-Van Hove (HVH) theorem \cite{hug} or the Bruckner theory \cite{bru64,Dab73}, nuclear symmetry energy \esym and its density slope $L(\rho) \equiv \left[3 \rho (\partial E_{\rm sym}/\partial \rho\right]_{\rho}$ have been expressed as \cite{XuC10,XuC11,Rchen}
\begin{eqnarray}
&& E_{\rm sym}(\rho) = \frac{1}{3} \frac{\hbar^2 k_F^2}{2 m_0^*} +
\frac{1}{2} U_{\rm sym}(\rho,k_{F}), \label{Esy2}
\\
&& L(\rho) \approx \frac{2}{3} \frac{\hbar^2 k_F^2}{2 m_0^*} +
\frac{3}{2} U_{\rm sym}(\rho,k_F) + \frac{dU_{\rm sym}}{dk}|_{k_F} k_F. \label{L2}
\end{eqnarray}
We emphasize here that the isoscalar effective mass $m_0^*$ enters explicitly the above expressions for both the magnitude and slope of the symmetry energy. Thus, there is no surprise at all that some transport model simulations have indicated that isospin-tracers and observables sensitive to the symmetry energy, such as the isospin diffusion and neutron/proton ratio of energetic nucleons in heavy-ion collisions, are also affected by the isoscalar nucleon effective mass. Moreover, the study of the isospin dependence of nucleon effective masses and the symmetry energy are intrinsically correlated by the same underlying interaction.
In fact, by neglecting the contributions of the momentum dependence of the isoscalar effective mass itself and the second-order symmetry (isoscalar $\delta^2$  term) potential, the neutron-proton effective mass splitting can be readily expressed in terms of the  $E_{\rm sym}(\rho_0)$ and $L(\rho_0)$ as \cite{LiBA13}
\begin{equation}
m^*_{n-p}(\rho_0,\delta)\approx\delta\cdot\frac{3E_{\rm sym}(\rho_0)-L(\rho_0)-\frac{1}{3}\frac{M}{M^*_0}E_{F}(\rho_0)}{E_F(\rho_0)(M/M_0^*)^2}. \label{npemass2}
\end{equation}
It is interesting to note that the $M_{\rm n}^*$ is equal, larger or smaller than the $M_{\rm p}^*$ depends on the symmetry energy and its slope. For example, using empirical values of \es0=31 MeV, $M_0^*/M=0.7$ and $E_F(\rho_0)=36$ MeV, a value of \l0$\leq 76$ MeV is required to get a positive $m^*_{n-p}(\rho_0,\delta).$ Interestingly, most of the extracted values of  $E_{\rm sym}(\rho_0)$ and $L(\rho_0)$ from both terrestrial experiments and astrophysical observations satisfy this condition \cite{LiBA13,EPJA}.

\section{Neutron-proton total effective mass splitting from nucleon-nucleus scatterings}\label{sec.III}
As illustrated in the examples shown in Figs. \ref{Muther} and \ref{Baldo}, nucleon effective masses are strongly density/momentum dependent, especially for the E-masses near the Fermi momenta. Ultimately, one has to find ways using, such as observables in heavy-ion collisions and/or neutron star observables, to probe the entire density/momentum dependences of all kinds of nucleon effective masses. Interestingly, nucleon-nucleus and electron-nucleus scattering data accumulated over several decades have already been used to constrain, respectively, the total effective mass and the E-mass at saturation density of nuclear matter. These provide important
boundaries for the nucleon effective masses and may be used already to constrain some of the models and the interactions.

\begin{figure}
\centerline{\includegraphics[width=9cm]{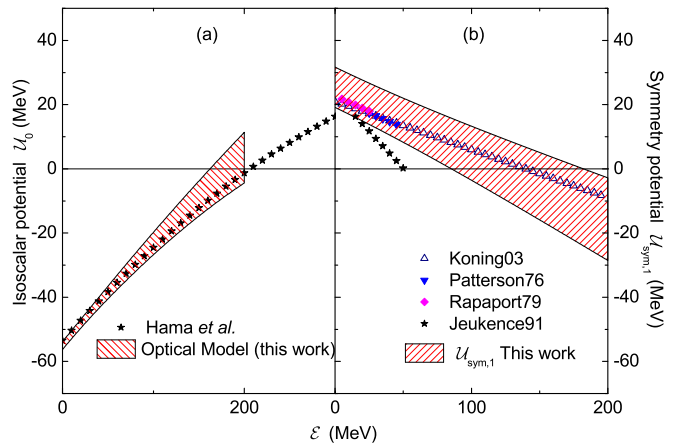}}
\label{Uoptical}
\caption{Energy dependent isoscalar $\mathcal{U}_0$(left) and isovector $\mathcal{U}_{sym}$ (right) nucleon potentials from analyzing nucleon-nucleus scattering data. Taken from Ref. \cite{XHLi}.}
\end{figure}

Optical model analyses of nucleon-nucleus scatterings have long been used to extract the momentum dependence of the isoscalar potential $dU_0/dk$ at saturation density and the associated nucleon isoscalar effective mass $m^*_0$ since the 1960s \cite{Hod94}. We summarize here the main findings of a global optical model analysis \cite{XHLi} of all 2249 data sets of reaction and angular differential cross sections of neutron and proton scattering on 234 targets at beam energies from 0.05 to 200 MeV available in the EXFOR database at the Brookhaven National Laboratory~\cite{Exfor}.
Shown on the left of Fig.\ 3 is a comparison of the nucleon isoscalar $\mathcal{U}_0$ potentials from this analysis (hatched bands) \cite{XHLi} and the Schr$\ddot{\mathrm{o}}$dinger equivalent isoscalar potential obtained earlier by Hama \textit{et al.}~\cite{Ham90}. Interestingly, they both give consistently an isoscalar effective mass of $m^{*}_0/m=0.65\pm 0.06$ consistent with its empirical value in the literature.
Shown on the right is the energy dependence of the nucleon isovector potential $\mathcal{U}_{sym}$ from several earlier studies~\cite{Kon03,Jeu91,Rap79,Pat76} and the most recent one (hatched bands) \cite{XHLi}.  Most of the earlier results are valid in low energy ranges. Albeit at different slopes,  they all clearly indicate a decreasing isovector optical potential with increase energy. After carefully translating the optical potentials into single-nucleon potentials in nuclear matter at saturation density, a neutron-proton total effective mass splitting $m^{*}_{n-p}=(0.41\pm0.15)\delta$ at saturation density was obtained \cite{XHLi}. Within its still large uncertainty range, it agrees with the BHF prediction of Ref. \cite{Baldo16} and the findings of Ref. \cite{LiBA13} that the total effective mass of neutrons is larger than that of protons in neutron-rich matter.  Moreover, a recent analysis of giant resonances in $^{208}$Pb~\cite{ZhangZ16} also found a value of $m^{*}_{n-p}$ consistent with this result. 

\section{Neutron-proton effective E-mass splitting from electron-nucleus scatterings}\label{sec.IV}
\begin{figure}[h!]
\centering
  \includegraphics[width=8.cm]{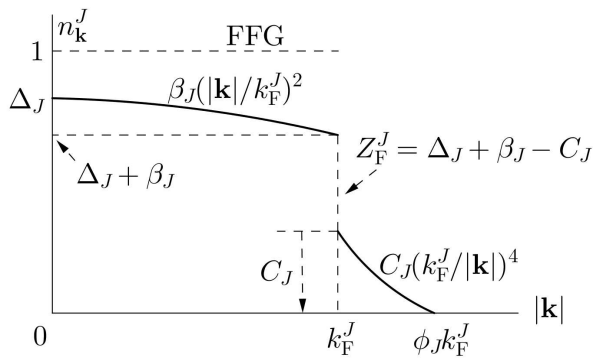}
  \caption{A sketch of the single nucleon momentum distribution with a high momentum tail above the Fermi surface. Taken from Ref. \cite{CaiLi16a}.}
  \label{mom-dis}
\end{figure}
The E-mass is related to the lifetime of quasi-particles in nuclear matter \cite{Jeu76,Neg81}.
Interestingly, the Migdal-Luttinger
theorem\,\cite{Mig57,Lut60} connects the nucleon E-mass directly with
the discontinuity $Z_{\rm{F}}^J\equiv
n_{\v{k}}^J(k^J_{\rm{F}-0})-n_{\v{k}}^J(k^J_{\rm{F}+0})$ of the
single-nucleon momentum distribution $n_{\v{k}}^J$ at the Fermi
momentum $k^J_{\rm{F}}$ illustrated in Fig. \ref{mom-dis} via
\begin{equation}\label{ML}
{M_{J}^{\ast,\rm{E}}}/{M}=1/Z_{\rm{F}}^J.
\end{equation}
In several recent studies \cite{CaiLi16a,CaiLi15,CaiLi16b},  a phenomenological $n_{\v{k}}^J$ of the form
\begin{equation}\label{MDGen}
n^J_{\v{k}}(\rho,\delta)=\left\{\begin{array}{ll}
\Delta_J+\beta_J{I}\left(\displaystyle{|\v{k}|}/{k_{\rm{F}}^J}\right),~~&0<|\v{k}|<k_{\rm{F}}^J,\\
&\\
\displaystyle{C}_J\left({k_{\rm{F}}^{J}}/{|\v{k}|}\right)^4,~~&k_{\rm{F}}^J<|\v{k}|<\phi_Jk_{\rm{F}}^J.
\end{array}\right.
\end{equation}
has been used. The $\Delta_J$ measures the
depletion of the Fermi sphere at zero momentum with respect to the
free Fermi gas (FFG) model prediction while the $\beta_J$ is the
strength of the momentum dependence $I(\v{k}/k_{\rm{F}}^J)$ of the
depletion near the Fermi surface. The parameters $\Delta_J$, $C_J$,
$\phi_J$ and $\beta_J$ depend linearly on $\delta$
according to $Y_J=Y_0(1+Y_1\tau_3^J\delta)$ as indicated by microscopic many-body theories \cite{mu04,Rio09,Yin13,Rio14}. The
amplitude ${C}_J$ and the cutoff coefficient $\phi_J$ determine the
fraction of nucleons above the Fermi surface. As discussed in detail in refs. \cite{CaiLi16a,CaiLi15,CaiLi16b},
except the $\beta_J$, all other parameters are constrained by experiments measuring the strength and isospin-dependence of
nuclear short range correlations (SRC) using electron-nucleus scattering at the Jefferson National Laboratory, see, e.g., \cite{Hen14,Hen15,Hen15b},
and applying Tan's adiabatic swipe theorem \cite{Tan08} to the equation of state of pure neutron matter calculated using the state-of-the-art microscopic theories \cite{Sch05,Epe09a,Gez10,Stew10,Tew13,Gez13}.

\begin{figure}[h!]
\centering
  \includegraphics[width=8.5cm]{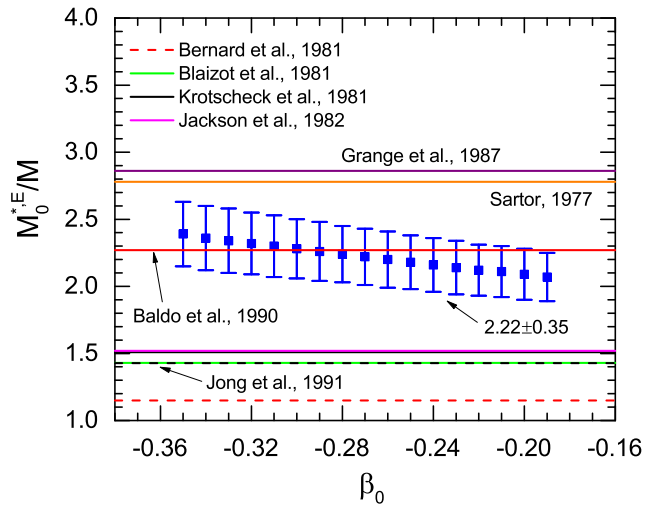}
  \caption{The nucleon effective E-mass in symmetric nuclear matter (blue lines with error bars) at normal density extracted from phenomenological nucleon momentum distribution constrained by electron-nucleus scattering data using the Migdal-Luttinger theorem in comparison with predictions of microscopic theories \cite{Ber81,Bla81,Kro81,Jac82,Gra87,Bal90,Sar77,Jon91} within the uncertainty range of the shape parameter $\beta_0$.
Taken from Ref.\cite{CaiLi16a}.}
  \label{Fig-Emass0}
\end{figure}

For symmetric nuclear matter (SNM), $M_0^{\ast,\rm{E}}/M\approx2.22\pm0.35$ was extracted using the Migdal-Luttinger theorem from the constrained phenomenological momentum distribution \cite{CaiLi16a}. Shown in Fig. \ref{Fig-Emass0} is a comparison of this value in  comparison with earlier predictions using (1) a semi-realistic parametrization through a relative
s-wave exponential nucleon-nucleon interaction potential (red dash line)\,\cite{Ber81}, (2) a Green's function method considering collective effects due to the coupling
of nucleons with the low-lying particle-hole excitations of the
medium (green solid line)\,\cite{Bla81}, (3) a correlated
basis function (CBF) method using the Reid and Bethe-Johnson
potentials (black and magenta solid lines)\,\cite{Kro81,Jac82}, (4)
two non-relativistic models with the Paris nuclear potential (purple and
red solid line)\,\cite{Gra87,Bal90}, (5) a low density expansion of
the optical potential (orange solid line)\,\cite{Sar77} and (6) a
relativistic Dirac-Brueckner approach (dash black
line)\,\cite{Jon91}, within the uncertain range of the $\beta_0$ parameter. It is
seen that the variation of $M_0^{\ast,\rm{E}}/M$ with $\beta_0$ is
rather small. Clearly, the predictions are rather diverse. The E-mass for SNM extracted from applying the Migdal-Luttinger theorem to the constrained phenomenological momentum distribution appears to be closer to the
BHF prediction by Baldo et al. \cite{Bal90}.

\begin{figure}[h!]
\centering
  \includegraphics[width=8.cm]{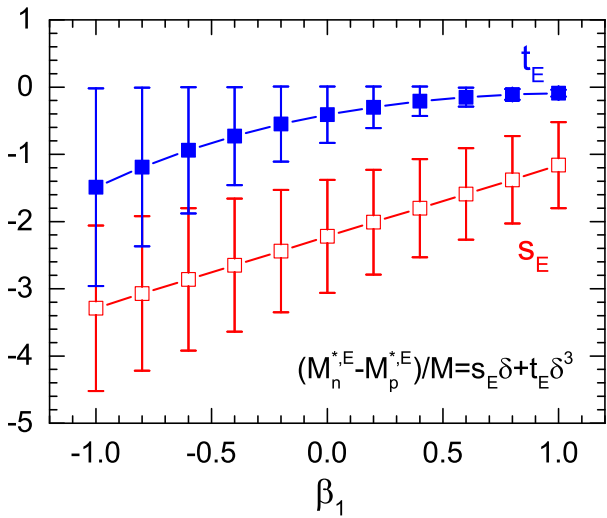}
  \caption{The linear and cubic splitting functions $s_{\rm{E}}$ and $t_{\rm{E}}$ at normal density within the uncertain range of the $\beta_1$-parameter characterizing the isospin-dependence of the nucleon momentum distribution near the Fermi surface. Taken from Ref.\cite{CaiLi16a}.}
  \label{Fig-Splitting}
\end{figure}

In isospin asymmetric matter, the neutron-proton E-mass splitting generally can be expanded in terms of $\delta$ as
\begin{equation}
\frac{M_{\rm{n}}^{\ast,\rm{E}}-M_{\rm{p}}^{\ast,\rm{E}}}{M}=s_{\rm{E}}\delta+t_{\rm{E}}\delta^3
+\mathcal{O}(\delta^5)
\end{equation}
where the $s_{\rm{E}}$ and $t_{\rm{E}}$ are shown in Fig. \ref{Fig-Splitting} within the uncertainty range of the $\beta_1$-parameter describing the isospin and momentum dependence of the nucleon Fermi surface.
It is interesting to note that the neutron E-mass is smaller than the proton E-mass, i.e.,
$M_{\rm{n}}^{\ast,\rm{E}}<M_{\rm{p}}^{\ast,\rm{E}}$ in neutron-rich medium. However, the neutron-proton E-mass splitting suffers from the large uncertainties due to the poorly known $\beta_1$-parameter. To improve the situation, one needs to
have more reliable knowledge about the isospin dependence of the nucleon momentum distribution around the Fermi surface. This information may be obtained from experiments
measuring the isospin dependence of the nucleon spectroscopic factors and the SRC strength in neutron-rich nuclei.

\section{Neutron-proton effective k-mass splitting and isospin dependence of nucleon mean free path in neutron-rich matter}\label{sec.V}
With the information about the total effective mass and E-mass, we can infer information about the k-mass from the relation of Eq. \ref{METK}.
In terms of the reduced mass, i.e., the dimensionless mass m ($M^*$ divided by $M$, etc), we have \cite{Cai16c}
\begin{eqnarray}
m_{\rm{n}}^{\ast,\rm{E}}&&\approx m_0^{\ast,\rm{E}}+\frac{1}{2}s_{\rm{E}}\delta,
m_{\rm{p}}^{\ast,\rm{E}}\approx m_0^{\ast,\rm{E}}-\frac{1}{2}s_{\rm{E}}\delta,\\\nonumber
&&m_{\rm{n}}^{\ast,\rm{E}}-m_{\rm{p}}^{\ast,\rm{E}}\approx
s_{\rm{E}}\delta,\\\nonumber
m_{\rm{n}}^{\ast,\rm{k}}&&\approx m_0^{\ast,\rm{k}}+\frac{1}{2}s_{\rm{k}}\delta,~~m_{\rm{p}}^{\ast,\rm{k}}\approx
m_0^{\ast,\rm{k}}-\frac{1}{2}s_{\rm{k}}\delta,\\\nonumber
&&m_{\rm{n}}^{\ast,\rm{k}}-m_{\rm{p}}^{\ast,\rm{k}}\approx
s_{\rm{k}}\delta.
\end{eqnarray}
Similarly, the linear splitting function $s$ for the total effective mass can be obtained from the nucleon-nucleus scattering data as discussed in Section \ref{sec.III}. The Eq. \ref{METK} then leads to
\begin{eqnarray}
s_{\rm{k}}&&=\frac{1}{m_0^{\ast,\rm{E}}}\left(s-\frac{s_{\rm{E}}m_0^{\ast}}{m_0^{\ast,\rm{E}}}\right)\approx0.50\pm0.24\\\nonumber
m_0^{\ast,\rm{k}}&&=\frac{m_0^{\ast}}{m_0^{\ast,\rm{E}}}\approx0.32\pm0.07
\end{eqnarray}
using $s\approx0.41\pm0.15,~~s_{\rm{E}}\approx-2.22\pm1.35,~~m_0^{\ast}\approx0.7\pm0.1$ and $m_0^{\rm{E}}\approx2.22\pm0.35.$
We notice that the value of $m_0^{\ast,\rm{k}}$ is significantly smaller than the empirical value of about 0.6 given in Refs. \cite{Jeu76,mu04}. The value of $s_k$ indicates that the k-mass has a significant isospin dependence compared to that of the total effective mass. The positive values of both $s$ and $s_k$ indicate that the total effective mass and the k-mass of neutrons (protons) increase (decrease) with the isospin asymmetry $\delta$,
while it is the opposite for the E-mass of nucleons.

The inferred values of the k-mass in SNM, its splitting for neutrons and protons and their dependence on the isospin asymmetry of the medium have interesting implications.
For example, the k-mass affects the nucleon MFP $\lambda$ and it was found necessary to explain qualitatively the observed large values of $\lambda$ for protons in SNM \,\cite{Neg81}.
It was emphasized that the space non-locality is as important as the Pauli blocking in determining the MFP.  In fact, it was shown quantitatively that estimates using the well-known expression 
$1/\lambda=\rho<\sigma>$ where $<\sigma>$ is the isospin averaged in-medium nucleon-nucleon cross section can not reproduce the experimental observations even if the Pauli blocking is considered unless the space non-locality through $M_k$ is also considered. More specifically,  the MFP $\lambda$ in nuclear medium is determined by \cite{Neg81}
\begin{equation}\label{MFP}
\lambda=\frac{k_{\rm{R}}}{2M_{\rm{k}}^{\ast}|W(E,k_{\rm{R}})|}
\end{equation}
where $k_{\rm{R}}=[2M(E-U(E,k_{\rm{R}}))]^{1/2}$ is the real part of the nucleon momentum and $W(E,k)$ is the
imaginary part of the potential. A reduced k-mass increases the nucleon MFP. Since the symmetry potential is repulsive (attractive) and the k-mass increases (decreases) for neutrons (protons), the $\frac{k_{\rm{R}}}{2M_{\rm{k}}^{\ast}}$ factor increases (decreases) for protons (neutrons) with the same energy as the isospin asymmetry $\delta$ increases. While existing analyses of nucleon-nucleus reaction data has not firmly established the isospin dependence of the imagine optical potential, recent many-body perturbation theory using chiral effective forces has clearly verified the Lane form of the imaginary potential \cite{Hol16}. Moreover, the magnitude of the isovecor part of the imaginary potential is appreciable compared to that of the real potential for nucleon energies below about 200 MeV. Obviously, whether protons have longer MFP than neutrons in neutron-rich medium depends on the isospin dependence of all three factors determining the $\lambda$ in Eq. \ref{MFP}. Previously, using a relativistic impulse approximation without considering the space non-locality it was found that neutrons have a longer MFP than protons with kinetic energies less than about 600 MeV, while it is the opposite at higher kinetic energies \cite{Jang}. In another study using a kinetic model considering the isospin dependence of both the Pauli blocking and nucleon-nucleon cross sections, but not the space-time non-locality necessary to reproduce the experimental observations \cite{Neg81,VJ}, it was found that neutrons always have longer MFP than protons \cite{Chen01}. More recently, using the definitions of  $1/\lambda_p=\rho_p\sigma_{pp}+\rho_n\sigma_{pn}$ and  $1/\lambda_n=\rho_n\sigma_{nn}+\rho_p\sigma_{np}$ with the in-medium nucleon-nucleon cross sections calculated within the DBHF approach, 
neutrons are found to have longer MFP than protons at kinetic energies less than about 300 MeV and the effect increases with the isospin asymmetry. At higher energies, however, neutrons and protons have approximately the same MFP \cite{Sama}. Thus, overall, we are still seeing rather model dependent predictions regarding the isospin and energy dependence of the nucleon MFP in neutron-rich matter while significant progress has been made in recent years. Further progress requires more complete knowledge about the isospin dependence of nucleon k-mass and its imaginary potential in neutron-rich matter. 

\section{Summary} \label{sec.VI}
In summary, due to the space-time non-locality of nuclear interactions single-nucleon potentials are momentum and/or energy dependent. Three distinct nucleon effective masses are normally used to
character the momentum/energy dependence of nucleon potentials. How do they depend on the density and isospin asymmetry of the medium? How are they different for neutrons and protons? These have been among the longstanding questions in nuclear physics. Answers to these questions have many interesting ramifications in both nuclear physics and astrophysics. In this talk, we briefly reviewed some of our recent efforts to answer these questions. In particular, we showed that the total effective mass $M^{\ast}_{J}$ for neutrons is higher than that for protons in neutron-rich matter at saturation density based on a comprehensive analysis of existing nucleon-nucleus scattering data. While the E-mass $M^{\ast,\rm{E}}_{J}$ for neutrons is less than that for protons in neutron-rich matter from applying the Migdal-Luttinger theorem to a phenomenological single-nucleon momentum distribution in nuclei constrained by recent electron-nucleus scattering experiments. Combining information about the isospin dependence of both the nucleon total effective mass and E-mass, we inferred the isospin dependence of nucleon k-mass. The latter is important for determining the nucleon MFP in neutron-rich matter. We also noticed some open questions regrading the nucleon effective masses to be further explored both theoretically and experimentally.

\section{Acknowledgement}
We would like to thank F.J. Fattoyev, W.J. Guo, O. Hen, W.G. Newton, E. Piasetzky and C. Xu for helpful discussions. 
This work was supported in part by the U.S. Department of Energy's Office of Science under Award Number DE-SC0013702; the CUSTIPEN (China-U.S. Theory Institute for Physics with Exotic Nuclei) under the U.S. Department of Energy Grant No. DE-SC0009971; the National Natural Science Foundation of China under grant no. 11320101004, 11275125, 11205083 and 11135011; the Major State Basic Research Development Program (973 Program) in China under Contract Nos. 2013CB834405 and 2015CB856904; the ``Shu Guang'' project supported by Shanghai Municipal Education Commission and Shanghai Education Development Foundation; the Program for Professor of Special Appointment (Eastern Scholar) at Shanghai Institutions of Higher Learning, the Science and Technology Commission of Shanghai Municipality (11DZ2260700); the construct program of the key discipline in Hunan province, the Research Foundation of Education Bureau of Hunan Province, China (Grant No.15A159); the Natural Science Foundation of Hunan Province, China (Grant No. 2015JJ3103) and the Innovation Group of Nuclear and Particle Physics in USC.

\end{document}